\documentclass[%
 reprint,
superscriptaddress,
nofootinbib,
 amsmath,amssymb,
prd,longbibliography,
floatfix,
]{revtex4-1}

\usepackage[T1]{fontenc}
\usepackage[utf8]{inputenc}
\usepackage{amsmath}
\usepackage{amsthm}
\usepackage{amsfonts,dsfont}
\usepackage{mathrsfs}
\usepackage{graphicx}

\usepackage{bbm}
\usepackage{hyperref}
\usepackage{color}
\usepackage{comment}

\usepackage{tikz}
\usetikzlibrary{positioning,shapes.geometric, arrows,calc,decorations.text}
\tikzstyle{arrow} = [->,>=stealth]
\tikzstyle{cir}=[inner sep=0,outer sep=0,circle,minimum height=0.1cm,draw=black]
\tikzstyle{blank}=[inner sep=0,outer sep=0,circle,minimum height=0.01cm,draw=black]

\theoremstyle{definition}

\newcommand\ket[1]{\left|#1\right>}
\newcommand\bra[1]{\left<#1\right|}

\newcommand\expect[1]{\left<#1\right>}

\newcommand\Ev{\mathbb{E}}
\newcommand\half{\frac{1}{2}}

\newcommand\ro{{\hat\rho}}
\newcommand\Ho{{\hat H}}
\newcommand\Vo{{\hat V}}
\newcommand\dens{{\varrho}}
\newcommand\denso{{\hat\varrho}}
\newcommand\Phio{{\hat\Phi}}

\newcommand\rv{\mathbf{r}}
\newcommand\sv{\mathbf{s}}
\newcommand\xv{\mathbf{x}}
\newcommand{\yv}[0]{\mathbf{y}}
\newcommand\si{\sigma}

\newcommand\Gscl{{\mathrm{Gscl}}}

\newcommand\Hcal{\mathcal{H}}
\newcommand\Ao{{\hat A}}
\newcommand\Bo{{\hat B}}

\newcommand{\upd}[0]{\mathrm{d}}

\definecolor{cbl}{rgb}{0,0,1}

\definecolor{crd}{rgb}{1,0,0}

\begin{document}
\title{Sourcing semiclassical gravity from spontaneously localized quantum matter}
\author{Antoine Tilloy}
\email{antoine.tilloy@ens.fr}
\affiliation{Laboratoire de Physique Th\'eorique, Ecole Normale Sup\'erieure de Paris, PSL, France}
\author{Lajos Di\'osi}
\email{diosi.lajos@wigner.mta.hu}
\affiliation{Wigner Research Center for Physics, H-1525 Budapest 114. P.O.Box 49, Hungary}

\date{\today}

\begin{abstract}
The possibility that a classical space-time and quantum matter cohabit at the deepest level, i.e. the possibility of having a fundamental and not phenomenological semiclassical gravity, is often disregarded for lack of a good candidate theory. The standard semiclassical theory suffers from fundamental inconsistencies (e.g.:  Schr\"odinger cat sources, faster-than-light communication and violation of the Born rule) which can only be ignored in simple typical situations. We harness the power of spontaneous localization models, historically constructed to solve the measurement problem in quantum mechanics, to build a consistent theory of (stochastic) semiclassical gravity in the Newtonian limit. Our model makes quantitative and potentially testable predictions: we recover the Newtonian pair potential up to a short distance cut-off (hence we predict no 1 particle self-interaction) and uncover an additional gravitational decoherence term which depends on the specifics of the underlying spontaneous localization model considered. 
We hint at a possible program to go past the Newtonian limit, towards a consistent general relativistic semiclassical gravity. 
\end{abstract}

\maketitle

\section{Introduction}
\label{intro}

Marrying quantum mechanics and gravity in the same physical theory is an extremely difficult endeavour, yet one that is desperately needed to understand the extreme scenarios where space-time becomes so singular that quantum effects should arise, e.g., in the early universe and in black holes. Insuring the consistency and unity of physics is  another no less important motivation. Many routes to quantum gravity, which aims at describing space-time as emerging from quantum dynamical degrees of freedom, have been explored \cite{kiefer2007}. These efforts have yielded interesting results but, after nearly half a century of hard work, none of them gave birth to an unanimously accepted quantum theory of gravity.

But do we really need to quantize gravity? Could we not imagine a dynamical classical curved space-time (in the sense of relativistic but not quantized) with some quantum matter in it? 
To build a complete semiclassical theory of gravity, one needs to say how quantum matter, described by state vectors $\ket{\Psi}$ in a Hilbert space, can source the curvature of a classical space-time. The standard approach to semiclassical gravity, due to M\o ller \cite{moller1962} and Rosenfeld \cite{rosenfeld1963}, is to use the quantum mechanical average to get a classical quantity out of the energy-momentum operator $\hat T_{ab}$ of quantized matter:
\begin{equation}\label{eq:rosenfeld}
G_{ab} = 8\pi G \bra{\Psi}\hat T_{ab}\ket{\Psi}
\end{equation}
where $G_{ab}$ is the Einstein tensor. So far, this is the only available
model of back-action of quantized matter on a classical space-time, indispensable 
to the description of our cosmology, stars and black holes.
When it is seen as an approximate theory, and as it ignores quantum fluctuations of $\hat T_{ab}$,
it becomes incorrect if the quantum state $\ket{\Psi}$ codes for large matter density
fluctuations \cite{page1981}. The semiclassiacal theory
possesses deep fundamental anomalies as well, the most spectacular one being faster-than-light 
communication \cite{eppley1977}. The standard semiclassical equation
\eqref{eq:rosenfeld} is consequently untenable as the building block of a fundamental theory. One faces Mielnik's deep-rooted alternative \cite{mielnik1974}:
\emph{either the gravitation is not classical or quantum mechanics
is not orthodox}. We are going to explore the latter option.

Interestingly, these anomalies already appear in the Newtonian regime as they are of quantum and not of relativistic nature. The defective prediction of the semiclassical theory in the case of large quantum  uncertainties of $\hat T_{ab}$ 
can be best understood in this simpler limit.  Consider
the superposition $\ket{\Psi}=\ket{A}+\ket{B}$ of a massive macroscopic object at two macroscopically different locations $A$ and $B$. Such Schr\"odinger cat states yield an intuitively and empirically incorrect source term in Eq.  \eqref{eq:rosenfeld} \cite{page1981}. 
The other anomaly \cite{eppley1977}, too, turns out to be basically quantum, i.e. essentially
unrelated to relativity and even to gravity. Having a quantum average $\langle\Ao\rangle$ in any deterministic dynamics spoils the linearity of quantum mechanics which allows faster-than-light communication \cite{gisin1990} and induces a break down of the statistical interpretation  (Born rule) \cite{mielnik1974,diosi1986}.

The objective of this article is to provide another way to source gravity from quantum matter in the non-relativistic limit, free of the inconsistencies of the standard approach. 
It turns out that all the previous anomalies can be dealt with through the use of spontaneous collapse models, a class of models initially aimed at solving the measurement problem in quantum mechanics.
 Most spontaneous wave function collapse models, see Sec. \ref{spontancollapse}, propose the
addition of a small non-linear stochastic term in the Schr\"odinger equation.
This small term is responsible for the dynamical collapse of macroscopic superpositions, i.e. the mechanism
\begin{equation}\label{collapse}
\ket{\Psi}=\ket{A}+\ket{B}~~\Rightarrow~~\ket{A}~\mathrm{or}~\ket{B}
\end{equation}
for macroscopic Schr\"odinger cat states,
only at the price of a negligible stochastic modification of the microscopic dynamics. 
These non-relativistic models are formally equivalent to the standard time-continuous
 quantum measurement 
(or monitoring) of the mass density operator 
$\denso=\hat T_{00}$ at each 
point in space by hidden unsharp detectors.
A crucial interest of this interpretation is that it naturally suggests to introduce the measured classical  
signal $T_{00}$ (time-continuous equivalent of von Neumann measurement outcomes), a now classical field which will be fluctuating around the quantum average value:
\begin{equation}\label{T00}
T_{00}=\langle\hat T_{00}\rangle+\delta T_{00},
\end{equation}
where $\delta T_{00}$ is the signal noise.

The following speculation, outlined already in \cite{diosi1990,diosi1998,diosi2011}, is now tempting.
Suppose we were able to construct a relativistic model of spontaneous collapse, 
i.e., formally monitoring the full energy-momentum tensor $\hat T_{ab}$. This would be useful in two ways.
First, the quantum monitoring of $\hat T_{ab}$ would suppress large quantum fluctuations of $\hat T_{ab}$, removing in particular the large Schr\"odinger cat ambiguities.
Second, the random signal $T_{ab}=\langle\hat T_{ab}\rangle+\delta T_{ab}$,  when used to
source the Einstein-tensor 
\begin{equation}\label{eq:einstein}
G_{ab} = 8\pi G \left(\langle\hat T_{ab}\rangle+\delta T_{ab}\right),
\end{equation}
would by construction respect the linear structure of quantum mechanics.
It would yield a back-action of quantized matter on space-time free of the 
anomalies of the deterministic semiclassical coupling \eqref{eq:rosenfeld}.
In terms of quantum control, back-action would be realized by a dynamical feedback 
conditioned on the signal.

Our objective with this theory of spontaneous-collapse-based stochastic semiclassical gravity is not to construct another approximate theory, more precise than the standard semiclassical one, to an exact yet unknown theory of quantum gravity. Rather, it is to propose a conceptually healthier semiclassical theory which is not plagued by foundational anomalies so it might in principle be the ultimate theory of gravity plus quantized matter. In this article, we will make this proposal precise and quantitative only in the non-relativistic sector. This is unfortunately needed for lack of a good relativistic spontaneous localization model. However, as the anomalies of the standard approach already show up in the Newtonian limit, showing that they can be cured represents a first promising step.

\textit{Outline--} The article is structured as follows. We first recall the standard possible approaches to semiclassical gravity in the Newtonian limit in Sec. \ref{semiclass}. We then give a short introduction to the spirit of spontaneous localization models and explain how they can be harnessed in semiclassical gravity in Sec. \ref{spontancollapse}. The core of our theory is then developed in mathematical details in Sec. \ref{monotoring_back-action}, first for a general spontaneous localization model, then in more details for the most studied ones, mainly Continuous Spontaneous Localization (CSL) and Di\' osi-Penrose (DP) models. Eventually we discuss some related works in Sec. \ref{related}, the main findings and their interpretation in the last section.

\section{Classical gravity vs quantized matter}
\label{semiclass}

Throughout all this article, we will consider $N$ particles of mass $m_1,...m_N$ evolving in 3 spatial dimensions.
We write $\Ho$ the many-body Hamiltonian (in the absence of gravity), $\ro=\ket{\Psi}\bra{\Psi}$ the many-body pure state density matrix, 
$\hat{\mathbf{x}}_n$ the position operator of the $n$-th particle and we use $\hbar=1$ through all sections but the appendix.

In the non-relativistic realm, where everything is technically easy, 
one can add gravity by a Newtonian pair-potential $\Vo_G$ in the Schr\"odinger-von Neumann equation:
\begin{equation}\label{vN}
\frac{\upd\ro}{\upd t}=-i[\Ho+\Vo_G,\ro],
\end{equation}
with the gravitational pair-potential:
\begin{equation}\label{Vo_G}
\Vo_G=-\frac{G}{2}\int\upd\rv\upd\sv\frac{\denso(\rv)\denso(\sv)}{\vert\rv-\sv\vert}
\end{equation}
where the spatial mass density of point-like constituents of masses $m_n$ and locations $\hat\xv_n$ is defined as\footnote{The results of our work are 
also valid for indistinguishable particles provided one takes
$\denso(\rv)=\sum_k m_k  \hat a_k^\dagger (\rv)\hat a_k(\rv)$
where $\hat a_k(\rv)$ and $\hat a_k^\dagger(\rv)$ are the local annihilation and creation operators of species of particle of mass $m_k$.}:
\begin{equation}\label{denso}
\denso(\rv)=\sum_n m_n\delta(\rv-\hat\xv_n).
\end{equation} 
Equations (\ref{vN}-\ref{denso}) represent the standard non-relativistic many-body quantum theory of gravitating constituents
and do not use the concept of gravitational field. This theory is free of any serious inconsistency (the infinite self-interaction energy has no dynamical consequence and is easily renormalized out) but it is not a very good candidate for a more general theory where the gravitational field is expected to be an autonomous entity. Admittedly, one can formally introduce the Newton field as a field operator slaved to the matter density: 
\begin{equation}\label{Phio}
\Phio(\rv)=-G\int\upd\sv\frac{\denso(\sv)}{\vert\rv-\sv\vert},
\end{equation}
but relating a classical gravitational field to this operator is a delicate issue and a central task of the present work.

Alternatively, and keeping an eye on general relativity where the gravitational field is a separate dynamical
entity interacting with matter, one  
studies matter-field interaction instead of the pair-potential $\Vo_G$.  The non-relativistic limit of the standard semiclassical approach \eqref{eq:rosenfeld} amounts to take a classical (i.e. not quantized) Newtonian potential $\Phi(\rv)$ satisfying the Poisson equation with the quantum average of the mass density operator as a source:
\begin{equation}\label{Poissonscl}
\nabla^2 \Phi(\rv)=4\pi G {\expect{\denso(\rv)}},
\end{equation}
yielding
\begin{equation}\label{Phiscl}
\Phi(\rv)=-G\int\upd\sv\frac{\expect{\denso(\sv)}}{\vert\rv-\sv\vert}
\end{equation}
which is a semiclassical counterpart of \eqref{Phio}.
In this setting,  the semiclassical Newton interaction can be introduced in the following way:
\begin{equation}\label{Vo_Gscl}
\Vo_{\Gscl}=\int\upd\rv\Phi(\rv)\denso(\rv),
\end{equation}
and then be used in the von Neumann equation \eqref{vN} in place of 
the pair-potential $\Vo_G$:
\begin{equation}\label{vNGscl}
\frac{\upd\ro}{\upd t}=-i[\Ho+\Vo_\Gscl,\ro].
\end{equation}
This equation, understood for pure states $\ro=\ket{\Psi}\bra{\Psi}$, is equivalent to the Schr\"odinger-Newton equation which has been
proposed earlier for the natural localization of quantum massive objects
\cite{diosi1984,penrose1996}:
\begin{equation}\label{SchN}
\frac{\upd\ket{\Psi}}{\upd t}=-i(\Ho+\Vo_\Gscl)\ket{\Psi}.
\end{equation}
This equation is the subject of intensive studies currently, cf., e.g.: \cite{bahrami2014,giulini2014,bekenstein2015,grossardt2015,grossardt2015effects}.  
The most salient feature of the Schr\"odinger-Newton equation is 
self-interaction: even the
single-body dynamics contains a gravitational self-interaction term.  Such a term would presumably not show up in a (so far elusive) theory of quantum gravity \cite{anastopoulos2014} but this does not discredit it a priori in the eventuality that space-time is fundamentally classical. 

However, and as natural as it may seem, the latter standard semiclassical approach is plagued by foundational problems.  
The most obvious one is that such a semiclassical coupling means non-linear deterministic quantum mechanics and, as we mentioned earlier,
this in itself leads to fatal anomalies.
The problem does not come from the way the gravitational potential is introduced in the Schr\"odinger von-Neumann equation, which is completely standard, but from the very way it is sourced from quantum matter.
The failure of this specific version of semiclassical gravity, at least when it is seen as a fundamental theory, is often taken as a strong argument in favour of the quantization of gravity. However, this only means that one of the most naive couplings between classical gravity and quantum matter does not do the trick. In the next section we discuss a way to get a stochastic classical mass density $\dens(\rv)$ from continuous localization models. This will give us
a consistent source of the classical Newton field $\Phi(\rv)$ that we will use in Sec. \ref{monotoring_back-action}. 
As we shall see, the proposed theory will solve the problems previously encountered and 
induce the Newtonian pair-potential from a classical gravitational field
without the inconsistencies arising from non-linearity.

\section{Spontaneous localization models}
\label{spontancollapse}
We now step back from gravity and review briefly a class of models originally aimed at solving the measurement problem in quantum mechanics.
A particular class of models, called spontaneous (or sometimes dynamical, objective) collapse (or sometimes localization, reduction) of the quantum state,   
describe the emergence of classical macroscopic phenomena dynamically and without reference to the presence of observers.
The standard unitary evolution of the quantum state
is modified by a universal weak 
collapse mechanism irrelevant
for microscopic degrees of freedom but which suppresses macroscopically large quantum uncertainties of the local
mass densities.
Spontaneous collapse theories, spearheaded by Ghirardi, Rimini, Weber (GRW) \cite{ghirardi1986} and Pearle \cite{pearle1989}, are reviewed in \cite{bassi2013}. The emphasis on mass density 
was proposed by one of the present authors \cite{diosi1987,diosi1989}. 

We will focus on the continuous versions of these models. They come 
in two main flavors, DP (for Di\'osi-Penrose) \cite{diosi1987,diosi1989,diosi1990,diosi2013,diosi2014,diosi2015,penrose1996,penrose1998,penrose2014} and CSL (for Continuous Spontaneous Localization) \cite{ghirardi1990,ghirardi1995}. 
Although they have some differences we will discuss later, these models share an important characteristic: their formalism can be interpreted as describing a quantum system subjected to a continuous monitoring of its (smeared) mass density, i.e. of the operators:
\begin{equation}\label{denso_si}
\denso_\si(\rv)=\sum_{n=0}^N m_n  g_\si(\rv-\hat\xv_n).
\end{equation}
The function $g_\sigma$
is a normalized Gaussian of width $\sigma$ and the smearing is necessary to keep the theory finite \cite{ghirardi1990gravity}. In what comes next we will generically define the \emph{smeared} version $f_\si$ of a field $f$ in the following way: $f_\si \equiv g_\si\ast f$. The analogy with continuous monitoring theory is only formal, the equations are the same but the interpretation is obviously different: in CSL and DP there is no detector and the spontaneous localization is taken as a fundamental fact of nature. The parallel is nevertheless extremely useful because it insures the consistency of the formalism: as they can be obtained from plain quantum theory, the equations of spontaneous localization models are guaranteed to preserve the statistical interpretation of the state vector. In what follows, we will sometimes use a vocabulary from continuous measurement theory (detectors, signal, etc.) but the reader should keep in mind that we only use it as a way to derive a consistent formalism, and not as if there were some real observer continuously doing measurements in nature.

We can illustrate this dual point of view on the example of the \emph{signal}. In continuous measurement theory, the signal is the time-continuous version of a measurement outcome and in the case we consider it reads:
\begin{equation}
\dens_t(\rv)=\langle\denso_\si\rangle_t +\delta\dens_t(\rv), 
\end{equation}
where $\delta\dens_t(\rv)$ is a white-noise in time with a potentially non trivial space correlator which we will specify later. In CSL or DP, the same quantity $\dens_t(\rv)$ is formally a fundamental stochastic (classical) field of the theory which can be taken as physical\footnote{More precisely, the signal can be taken as the primitive ontology \cite{bell1976,allori2008,allori2014} of CSL and DP, i.e. as the only local physical stuff living in space, a point of view which was advocated in \cite{diosi2015}. We shall not develop this idea in details here but this is another way to see why one would naturally want to use the signal to source gravity.}.
In the following section, we will use this field to source gravity. Why is this a better idea than simply using the average $\langle\denso_\si\rangle_t$ as one would do in standard semiclassical gravity? The answer is that the signal is simply a measurement result, and modifying the subsequent evolution of a quantum system based on a measurement result is allowed in orthodox quantum theory! Therefore, if we use the stochastic field $\dens_t(\rv)$ to source gravity, formally as in a feedback scheme, we will get a theory which, by construction, will be free of the anomalies of standard semiclassical gravity.

In the next section we will apply this program and marry spontaneous collapse with (Newtonian) semiclassical gravity, first for a generic spontaneous collapse model and then in more details in the specific cases of CSL and DP. 
Technically, we rely on the density matrix formalism and stochastic master equations. Nevertheless everything can be recast in the state vector formalism, with the corresponding
stochastic  Schr\"odinger equations.

\section{Spontaneous monitoring of mass density, and back-action on gravity}
\label{monotoring_back-action}
\subsection{General case}
\begin{figure}
\includegraphics[width=.9\columnwidth,trim = 0cm 23.5cm 12cm 0cm, clip]{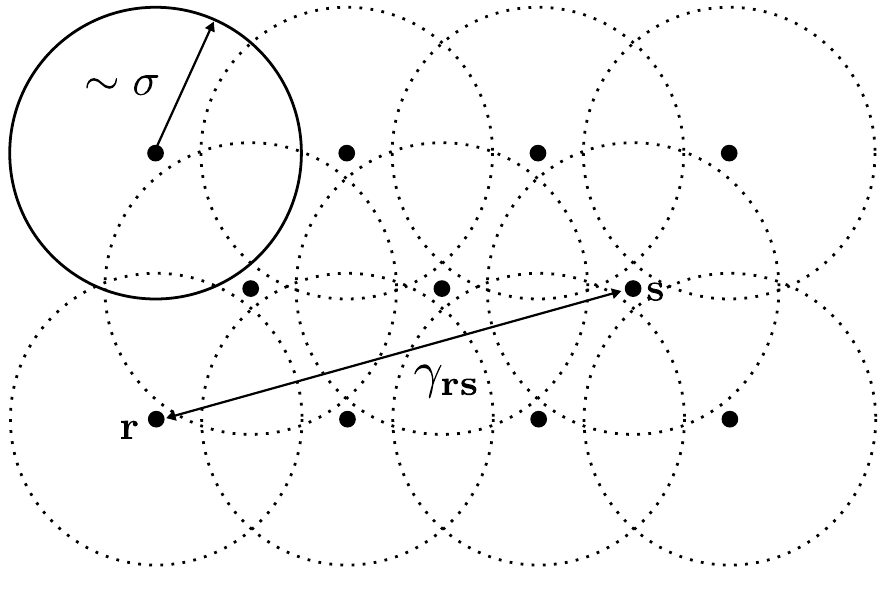}
\caption{Intuitive representation of the detectors (here in 2D). The locations of the detectors are represented by black dots on a lattice for simplicity but the reader should imagine that they fill space continuously (or that the lattice spacing is much smaller than all the other length scales). The radius $\sigma$ represents the spatial resolution of a single detector and $\gamma_{\rv\sv}$ codes for the correlation of the outputs of a detector in $\rv$ and one in $\sv$. Notice that $\sigma$ and $\gamma_{\rv\sv}$ represent completely different physical quantities and as such, they can be chosen independently.}
\label{fig:detectors}
\end{figure}

We now consider a general many-particle spontaneous localization model which includes CSL and DP as specific cases. Formally it is equivalent to the continuous monitoring of the mass density by (hidden) detectors of spatial resolution $\sigma$. The detectors are also possibly entangled, which correlates their measurement outcomes. As we have claimed in the previous section, the continuous equivalent of a von-Neumann measurement result, called the signal, reads in this context:
\begin{equation}\label{signaldens}
\dens_t(\rv)=\langle\denso_\si\rangle_t +\delta\dens_t(\rv),
\end{equation}
where $\delta\dens_t(\rv)$ is a spatially correlated white-noise (understood with the It\^o convention):
\begin{equation}\label{corrdens}
\Ev[\delta\dens_t(\rv) \delta\dens_\tau(\sv)]=\gamma^{-1}_{\rv\sv}\delta(t-\tau),
\end{equation}
and $\gamma_{\rv\sv}$ is a non-negative kernel which intuitively encodes the correlation between the detectors at positions $\rv$ and $\sv$ (see Fig. \ref{fig:detectors}). 
The stochastic master equation (SME) prescribing the dynamics of the system density matrix reads:
\begin{eqnarray}\label{SMEcoll}
\frac{\upd\ro}{\upd t}&=&-i[\Ho,\ro]-\int \upd\rv\upd\sv\frac{\gamma_{\rv\sv}}{8}[\denso_\si(\rv),[\denso_\si(\sv),\ro]]\nonumber\\
                            & & +\int \upd\rv\upd\sv\frac{\gamma_{\rv\sv}}{2}\Hcal[\denso_\si(\rv)]\ro \delta\dens(\sv),
\end{eqnarray}
where 
\begin{equation}\label{defHcal}
\Hcal[\denso_\si(\rv)](\hat\rho)=\left\{\denso_\si(\rv)-\expect{\denso_\si(\rv)}_t,\ro_t \right\}.
\end{equation}
The Eqs. (\ref{signaldens}-\ref{defHcal}) complete the specification of our many-particle spontaneous localization model without gravity. 
We give more mathematical 
details on the derivation of these equations from continuous measurement theory in the appendix. In what follows, we will take them as given but we can nevertheless give the reader a quick heuristic understanding of the different terms. The deterministic term with the double commutator in \eqref{SMEcoll} implements the decoherence induced by the coupling with the detectors and tends to reduce the state purity and to make the density matrix diagonal in the position basis of bulky objects. The stochastic term implements the localization coming from the conditioning on the measurement results: it drives the density matrix towards localized states and exactly compensate the purity loss induced by the decoherence term. For a typical spontaneous localization model like CSL or DP, these two terms are small in the sense that they have a negligible impact on the dynamics of microscopic systems but dominate for macroscopic systems which become well localized in position.

We are now in the position to construct the back-action of the quantized matter on the classical gravitational field. Technically, we make the ``monitored'' value  $\dens_t(\rv)$ \eqref{signaldens} of the matter density the source of the classical
 Newton potential in the Poisson eq. \eqref{Poissonscl}, instead of $\expect{\denso(\rv)}$:
 \begin{equation}
 \nabla^2 \Phi(\rv)=4\pi G {\dens(\rv)}.
 \end{equation}
We should mention that this equation was already present in \cite{pearle1996} but the objective of the authors was very different: their aim was to find a possible gravitational origin for the white noise in spontaneous localization models, our objective is essentially opposite. 
 The modified semiclassical Newton potential then becomes stochastic and 
 takes the form:
 \begin{equation}\label{Phist}
\Phi(\rv)=-G\int\upd\sv\frac{\dens(\sv)}{\vert\rv-\sv\vert}.
\end{equation}
Inserting this field, we obtain the stochastic semiclassical interaction $\Vo_\Gscl$ of the same form \eqref{Vo_Gscl} as before only with the small technical difference that we leave an option of $\si$-smearing open:
\begin{eqnarray}\label{Vo_Gsclst}
\Vo_\Gscl&=&\int\upd\rv\Phi(\rv)\denso_{(\si)}(\rv)\nonumber\\
                   &=&\int\upd\rv\dens(\rv)\Phio_{(\si)}(\rv),
\end{eqnarray}
where $(\si)$ denotes an optional convolution with $g_\si$. As we will see, this optional smearing will be necessary for DP to avoid divergences but superfluous for CSL.

To introduce this potential into the evolution, we have to be careful because of the multiplicative white-noise (again in the It\^o convention). Technically, we introduce it as if it were a feedback, i.e. have the potential act an infinitesimal amount of time after the ``free'' evolution given by Eq. \eqref{SMEcoll}:
\begin{equation}\label{fb}
\ro+\upd\ro=e^{-i\Vo_{\Gscl} \upd t}(\ro+\upd\ro^{\text{free}})e^{i\Vo_{\Gscl} \upd t}.
\end{equation}
More details on how to implement a generic feedback scheme in continuous measurement theory are provided in the appendix \ref{monitoring_feedback}. Expanding the exponential up to second order then gives the SME for the complete evolution:
\begin{eqnarray}\label{SMEfbcoll}
\frac{\upd\ro}{\upd t}&=&-i\left[\Ho+\Vo_{G,\si}+\int \upd\rv\delta\dens(\rv)\Phio_{(\si)},\ro\right]\nonumber\\
& &\hskip-24pt-\!\!\!\int \!\!\!\upd\rv\upd\sv\!\left(\!\!\frac{\gamma_{\rv\sv}}{8}[\denso_\si(\rv),[\denso_\si(\sv),\ro]]
                                          \!+\!\frac{\gamma_{\rv\sv}^{-1}}{2}[\Phio_{(\si)}(\rv),[\Phio_{(\si)}(\sv),\ro]]\!\!\right)\nonumber\\
                            & & +\int \upd\rv\upd\sv\frac{\gamma_{\rv\sv}}{2}\Hcal[\denso_\si(\rv)]\ro \delta\dens(\sv),
\end{eqnarray}
where the deterministic part of the back-action Hamiltonian yields
\begin{equation}\label{Vback}
\Vo_{G,\si}=\half\int \upd\rv \denso_\si(\rv)\Phio_{(\si)}(\rv).
\end{equation}
This is a remarkable result, it is independent of the strength $\gamma_{\rv\sv}$ we assume for
monitoring the mass density $\denso_\si(\rv)$. We can write it in the equivalent form:
\begin{equation}\label{Vback1}
\Vo_{G,\si}=-\frac{G}{2}\int \upd\rv\upd\sv \frac{\denso_\si(\rv)\denso_{(\si)}(\sv)}{\vert\rv-\sv\vert}.
\end{equation}
This is  the Newton pair-potential of the standard gravitational interaction 
\eqref{Vo_G} --- up to smearing of the mass density around the point-like constituents. Note that the semiclassical self-interaction of individual
constituents, one of the characteristic features of deterministic standard semiclassical gravity, has been cancelled in our signal-based stochastic semiclassical gravity. More precisely, self interaction only shifts all the energies by a finite quantity (diverging when $\sigma\rightarrow 0$) and thus has no dynamical consequence.
 
Let us summarize our model of stochastic semiclassical gravity.
The mass density $\denso_\si$ is spontaneously monitored yielding
the signal $\dens$ \eqref{signaldens} containing the white noise $\delta\dens$
\eqref{corrdens}, which is used to create the back-action \emph{on} gravity. 
The quantum state $\ro$ is evolved by the SME \eqref{SMEfbcoll} which implements the back-action \emph{of} gravity.
Monitoring leads to local decoherence in $\denso_\si$. Gravitational back-action
leads to an additional  local decoherence in $\Phio_{(\si)}$. 
Back-action generates the standard Newton pair-potential
up to a microscopic smearing. The stochastic Hamiltonian term
directly corresponds to the back-action of the signal noise $\delta\dens$.
Eventually, the non-Hamiltonian stochastic term is responsible for the time-continuous
collapse, i.e., localization in $\denso_\si$, preventing large quantum
fluctuations of mass density
and Schr\"odinger cat sources in particular.

\subsection{CSL: Continuous spontaneous localization}
\label{CSL}
This is the simplest spontaneous localization model one can think of. It is a bit ad-hoc in the sense that it is historically motivated only by the resolution of the measurement problem and has two free parameters, unrelated to gravity. It is on the other hand very simple. It is formally equivalent to the continuous monitoring of the mass density by independent (hidden) detectors, i.e. it uses a trivial correlator:
\begin{equation}\label{gammaCSL}
\gamma_{\rv\sv}=\gamma\delta(\rv-\sv).
\end{equation}
The standard choice of GRW, which is compatible with experiments carried out up to now, is to take a space cutoff $\sigma\sim10^{-5}$cm and a strength parameter $\gamma\sim\hbar^2\times10^{16}\mathrm{cm}^3\mathrm{g}^{-2}\mathrm{s}^{-1}$. Other choices are possible, see for example Adler \cite{adler2007}, but combinations of significantly larger $\gamma$ and significantly smaller $\sigma$, yielding fast spontaneous collapse, are excluded by experiments \cite{feldmann2012}.

Let us recall, for completeness, the measured (monitored) value \eqref{signaldens} of $\denso_\si(\rv)$: 
\begin{equation}\label{densCSL}
\dens_{t}(\rv)={\expect{\denso_\si(\rv)}}_t + \delta\dens_t(\rv),
\end{equation}
where, according to \eqref{corrdens} and \eqref{gammaCSL}:
\begin{equation}\label{corrdensCSL}
\Ev [\delta\dens_t(\rv)\delta\dens_\tau(\sv)]=\frac{1}{\gamma}\delta(t-\tau)\delta(\rv-\sv).
\end{equation}
In CSL it is possible to define the gravitational potential \eqref{Vo_Gsclst} via the sharp density without getting infinities so we do it for simplicity:
\begin{equation}\label{Vo_GsclstCSL}
\Vo_\Gscl=\int\upd\rv\Phi(\rv)\denso(\rv).
\end{equation}
This yields the SME \eqref{SMEfbcoll}:
\begin{eqnarray}\label{SMEfbCSL}
\frac{\upd\ro}{\upd t}&=&-i\left[\Ho+\Vo_{G,\si}+\int \upd\rv\delta\dens(\rv)\Phio,\ro\right]\nonumber\\
& &\hskip-24pt-\int \upd\rv \left(\frac{\gamma}{8}[\denso_\si(\rv),[\denso_\si(\rv),\ro]]
                                          +\frac{1}{2\gamma}[\Phio(\rv),[\Phio(\rv),\ro]]\right)\nonumber\\
                            & & +\int \upd\rv \frac{\gamma}{2}\Hcal[\denso_\si(\rv)]\ro \delta\dens(\sv),
\end{eqnarray}
where the back-action Hamiltonian \eqref{Vback} takes the following
symmetric form:
\begin{equation}\label{VbackCSL}
\Vo_{G,\si}=-\frac{G}{2}\int \upd\rv\upd\sv \frac{\denso_{\si/\sqrt{2}}(\rv)\denso_{\si/\sqrt{2}}(\sv)}{\vert\rv-\sv\vert}.
\end{equation}
The first decoherence term $-\int \upd\rv\frac{\gamma}{8}[\denso_\si(\rv),[\denso_\si(\rv),\ro]]$ is already present in CSL so we will only discuss the second one, $\hat D_G[\hat \rho]=-\int \upd\rv \frac{1}{2\gamma}[\Phio(\rv),[\Phio(\rv),\ro]]$, introduced by the back-action noise of the coupling with gravity. We consider the case of a single particle of mass $m$ and density matrix $\rho(\xv,\yv)$. It this case,  the contribution of the back-action decoherence
to the  dynamics of $\rho(\xv,\yv)$ can luckily be computed explicitly:
\begin{equation}
\begin{split}
{D_G[\hat \rho](\xv,\yv)}&=-\frac{G^2 m^2}{8\gamma}\!\int \!\!\upd\rv\!\left(\frac{1}{|\rv-\xv|}-\frac{1}{|\rv-\yv|}\right)^2\!\!\rho(\xv,\yv)\\
&=-\frac{\pi G^2m^2}{2\gamma} |\xv-\yv|\; \rho(\xv,\yv)
\end{split}
\end{equation}
So the back-action decoherence term simply damps the phases of the density matrix proportionally with the distance $|\xv-\yv|$ separating the positions (in real space) considered. This is to be contrasted with decoherence coming from CSL's $\denso_\si$-decoherence itself  
 which increases quadratically for short ($|\xv-\yv| \ll \sigma$) and saturates for long ($|\xv-\yv|\gg \sigma$) distances. This means that depending on the value of $\gamma$, either back-action decoherence dominates at every scale or it dominates at very short and very long distances. Back-action decoherence also globally increases when the strength of collapse $\gamma$ decreases and could give a stringent lower bound on the collapse rate. This is fundamental because the only lower bounds currently available for the collapse rate are of metaphysical origin \cite{feldmann2012}, i.e. one requires that the collapse model gives a philosophically satisfactory description of the macroscopic world. The coupling with gravity thus provides empirical constraints on the lower bound although the details deserve additional investigations.

\subsection{DP --- Gravity related spontaneous collapse}
\label{DP}
This model was historically constructed in order to reduce macroscopic quantum fluctuations of mass density \cite{diosi1987,diosi1987} and to solve the measurement problem with the help of heuristic considerations involving gravity which we will not develop here. It is equivalent with spontaneous monitoring of $\denso_\si(\rv)$  by spatially correlated (hidden) detectors. DP thus uses the slightly less trivial correlator:
\begin{equation}\label{gammaDP}
\gamma_{\rv\sv}=\kappa G\frac{1}{\vert\rv-\sv\vert},
\end{equation}
where $G$ is the gravitational constant.
The obvious interest of this form is that the constant $\kappa$ is now a dimensionless parameter which will be fixed to $2$ soon by an additional physical consideration. In DP, the localization strength is thus tightly related to gravity right from the start and there is one less free parameter than in CSL.
\footnote{There is however a difficulty to fix the spacial cut-off $\sigma$ in a way which is consistent with experiments \cite{bahrami2014DP}. In \cite{diosi2014}, one of the present authors nevertheless supported the natural cutoff $\si\sim10^{-12}$cm and worked out an SME to lift earlier conflicts with experiments.}. The more complicated form of \eqref{gammaDP} will also help us get a more symmetric SME in the end.

The inverse kernel is quasi-local:
\begin{equation}\label{gammainvDP}
\gamma_{\rv\sv}^{-1}=-\frac{1}{4\pi\kappa G}\delta(t-\tau)\nabla^2\delta(\rv-\sv).
\end{equation}
Recall the measured (monitored) value \eqref{signaldens} of $\denso_\si(\rv)$, for completeness: 
\begin{equation}\label{densDP}
\dens_{\si,t}(\rv)={\expect{\denso_\si(\rv)}}_t + \delta\dens_t(\rv),
\end{equation}
where, according to \eqref{corrdens} and \eqref{gammainvDP}:
\begin{equation}\label{corrdensDP}
\Ev [\delta\dens_t(\rv)\delta\dens_\tau(\sv)]=-\frac{1}{4\pi\kappa G}\delta(t-\tau)\nabla^2\delta(\rv-\sv).
\end{equation}
In DP we have to use the smeared density in \eqref{Vo_Gsclst} to avoid divergences:
\begin{equation}\label{Vo_GsclstDP}
\Vo_\Gscl=\int\upd\rv\Phi(\rv)\denso_\si(\rv).
\end{equation}
This yields the SME \eqref{SMEfbcoll}:
\begin{eqnarray}\label{SMEfbDP}
\frac{\upd\ro}{\upd t}&=&-i\left[\Ho+\Vo_{G,\si}+\int \upd\rv\delta\dens(\rv)\Phio_\si,\ro\right]\nonumber\\
                            & &-\frac{\kappa G}{8}\int \frac{\upd\rv\upd\sv}{\vert\rv-\sv\vert}[\denso_\si(\rv),[\denso_\si(\rv),\ro]]\nonumber\\
                            & &-\frac{1}{8\pi\kappa G}\int \upd\rv [\nabla\Phio_\si(\rv),[\nabla\Phio_\si(\rv),\ro]]\nonumber\\
                            & & +\frac{\kappa G}{2}\int\frac{\upd\rv\upd\sv}{{\vert\rv-\sv\vert}}\Hcal[\denso_\si(\rv)]\ro \delta\dens(\sv),
\end{eqnarray}
where the back-action Hamiltonian \eqref{Vback} takes the following
symmetric form:
\begin{equation}\label{VbackDP}
\Vo_{G,\si}=-\frac{G}{2}\int \upd\rv\upd\sv \frac{\denso_{\si}(\rv)\denso_{\si}(\sv)}{\vert\rv-\sv\vert}.
\end{equation}

Observe that, due to the quasi-locality of  \eqref{gammainvDP}, the back-action decoherence has become local in the Newton acceleration field $\nabla\Phi_\si$.
Interestingly, this structure coincides with the typical non-local $\denso_\si$-decoherence term caused by DP spontaneous collapses. The two decoherence terms can be united into the following local form: 
\begin{eqnarray}\label{decohDP}
& &-\frac{\kappa G}{8}\int \frac{\upd\rv\upd\sv}{\vert\rv-\sv\vert}[\denso_\si(\rv),[\denso_\si(\rv),\ro]]\nonumber\\
& &-\frac{1}{8\pi\kappa G}\int \upd\rv [\nabla\Phio_\si(\rv),[\nabla\Phio_\si(\rv),\ro]]\nonumber\\
&=&-\left(\frac{\kappa}{4}+\frac{1}{\kappa}\right)\frac{1}{8\pi G} \int \upd\rv [\nabla\Phio_\si(\rv),[\nabla\Phio_\si(\rv),\ro]].
\end{eqnarray}
If we now require that decoherence be minimal in the full model of DP-based stochastic semiclassical gravity, which seems to be a reasonable physical assumption, we get $\kappa=2$.
The ultimate local form of the SME \eqref{SMEfbDP} then
reads:
\begin{eqnarray}\label{SMEfbDPloc}
\frac{\upd\ro}{\upd t}&=&-i\left[\Ho+\Vo_{G,\si}+\int \upd\rv\delta\dens(\rv)\Phio_\si,\ro\right]\nonumber\\
                            & &-\frac{1}{8\pi G}\int \upd\rv [\nabla\Phio_\si(\rv),[\nabla\Phio_\si(\rv),\ro]]\nonumber\\
                            & & -\int \upd\rv \Hcal[\Phio_\si(\rv)]\ro \delta\dens(\rv).
\end{eqnarray}
where back-action has just doubled the decoherence term of the initial DP model. Historically,  this doubling
had been derived in \cite{diosi1990} while \cite{diosi2011} had cancelled it by an ad-hoc mean-field ansatz.

\section{Related approaches}\label{related}
This work can be contrasted with earlier non-standard approaches to semiclassical gravity.
A stochastic semiclassical theory of gravity \cite{hu1994} (see also \cite{martin1999}) was proposed a long time ago. The objective of the authors was to phenomenologically relax the ignorance of the standard source term $\bra{\Psi}\hat T_{ab}\ket{\Psi}$ with respect to quantum fluctuations.  
In their model, the Einstein equation takes a stochastic form similar to ours \eqref{eq:einstein} but  the stochastic noise $\delta T_{ab}$
is constructed  to mimic the quantum fluctuations of  $\hat T_{ab}$ and is not related to spontaneous monitoring. 
The dynamics of the quantum state $\ket{\Psi}$ remains non-linear on each sample of the classical background space-time and does not include any collapse mechanism. As a result, it still suffers from the anomalies of the standard semiclassical theory which are caused by the coupling to $\bra{\Psi}\hat T_{ab}\ket{\Psi}$. Similarly, even if the recent approach \cite{derakhshani2014} includes the GRW 
discrete spontaneous collapse mechanism which suppresses macroscopic Schr\"odinger-cat states, the author uses the average mass density to source the gravitational field which pushes him to propose a statistical interpretation different from the standard Born rule.

Recently, the Newtonian sector of semiclassical gravity has been investigated with approaches bearing some similarities with the present work. In \cite{nimmrichter2015}, the authors attack a slightly more specific problem and attempt to cure the faster-than-light communication anomaly of
the  Schr\"odinger-Newton equation \eqref{SchN}. They supplement it with an ad-hoc nonlinear stochastic term.
The modified dynamics  leads, like ours does, to the unconditional DP master equation which is free of the usual anomalies of semiclassical gravity.
Yet, interestingly, the semiclassical coupling cancels out, without leaving  the Newton pair-potential behind.  This ad-hoc stochastic semiclassical gravity thus lacks a gravitational interaction. 
On the contrary, in our approach, the analogue of the Schr\"odinger-Newton equation \eqref{SchN} reads:
\begin{eqnarray}\label{SSchN}
\frac{\upd\ket{\Psi}}{\upd t}&=&-i\left(\Ho+\Vo_{G,\si}\right)\ket{\Psi}\\
                            & &-\frac{1}{8\pi G}\int \upd\rv \left(\nabla\Phio_\si(\rv)-\expect{\nabla\Phio_\si(\rv)}\right)^2\ket{\Psi}\nonumber\\
                            & & -(1+i)\int \upd\rv \left(\Phio_\si(\rv)-\expect{\Phio_\si(\rv)}\right)\delta\dens(\rv)\ket{\Psi}\nonumber
\end{eqnarray}
which can be obtained from \eqref{SMEfbDPloc} writing $\ro=\ket{\Psi}\bra{\Psi}$.
Note that the semiclassical potential of \eqref{SchN} becomes a pair-potential in our version.

We should eventually mention the pioneering work of Kafri \textit{et al.} \cite{kafri2014}. The authors
formulate a theory conceptually similar to ours in a quantum communication context. Classical (non-quantum) gravity between
two objects is interpreted as a classical measurement channel. In this context, the word ``classical'' means the channel cannot entangle the two separated objects. The two-body toy model of \cite{kafri2014} turns out to be a specific case of the stochastic semiclassical gravity with the DP signal (Sec. \ref{DP}) and  anticipates its remarkable quantum informational features. The toy model operates at the
noise threshold where the Newton interaction cannot entangle the two objects any more. How our theory satisfies this informational condition of classicality in the general case is an interesting subject for future work.  
 
\section{Summary, outlook}

In this article, we have shown how to source a classical gravitational field from spontaneously localized quantum particles. Using the fact that localization models are formally equivalent to continuous quantum measurement models, we have introduced a new quantity, the signal, and promoted it to the status of physical source of the gravitational field. In terms of quantum control, back-action has been formally realized as a dynamical feedback based on the signal. One of us argued earlier that signal, i.e.: quantum measurement
outcome, is the only variable tangible for control like feedback \cite{diosi2012}. 
This fact gives a justification for spontaneous collapse models as they seem to be the \emph{only} way to couple a quantum theory of matter with a classical theory of space-time.
Spontaneous localization models are seen by many as an ad-hoc method to solve the measurement problem. However, the fact that they seem unavoidable for a classical-quantum coupling is one of their less known yet remarkable feature.

Our model is mathematically tractable and makes some precise and testable predictions. We have shown that it gives rise to the expected Newtonian pair potential up to a small correction at short distances coming from the spatial cut-off of the underlying localization model. This pair potential does not give rise to 1 particle self-interaction which means that such a self interaction is not a necessary consequence of semiclassical gravity as it is often believed. Failing to see self-interaction in experiments \cite{grossardt2015,grossardt2015effects} would prove the failure of the Schr\"odinger-Newton equation, but it would not imply the quantization of gravity. Additionally, gravitational back-action introduces a new decoherence term which depends strongly on the underlying localization model chosen. In CSL, gravitational back-action adds a decoherence term which increases linearly with distance, has consequently no characteristic scale and is independent on the microscopic details of the theory. This decoherence term also globally increases when the collapse strength decreases which makes low values of $\gamma$ experimentally falsifiable (and not only metaphysically unsatisfying). In DP, the additional decoherence term takes the same form as the original intrinsic decoherence which makes the final equations very symmetric. In this case, the fact that gravitational decoherence increases when the collapse strength decreases allows us to find a global 
minimum for decoherence which singles out the gravitational constant $G$ as the collapse strength.

Eventually, as it solves the inconsistencies of standard semiclassical gravity in the Newtonian regime, our model --or class of models-- is a sound first step in the construction of a full relativistic semiclassical theory of gravity.
In the relativistic realm, the covariant equation of back-action should be given by the stochastic Einstein equation
\eqref{eq:einstein} with noise $\delta T_{ab}$. However even the basic principles governing a theory of convariant continuous monitoring are still problematic.
Working with white-noises (the Markovian case),
the only Lorentz covariant possibility is the one with both temporally and spatially
uncorrelated noise, which leads to fatal divergences \cite{Pearle1990} once it is coupled to local quantum fields. 
The divergences can be eliminated through a covariant smearing depending on $\expect{T_{ab}}$ \cite{bedingham2011}, but such a non-linear addition would be in conflict with the aims of the present article.
Another possibility is to use colored noises right from start.
Such a non-Markovian field-theoretic formalism of monitoring and feedback was laid down a long time ago \cite{diosi1990}.  
In the Markovian limit, it yielded --albeit in a different formalism-- exactly the theory presented in the Appendix and used throughout the present article. 
However a serious difficulty of this approach comes from the non-instantaneous availability of the non-Markovian signal \cite{wiseman2008,diosi2008,diosi2008e}. Making our model relativistic is thus not just a purely technical task and many additional unexpected obstacles might be encountered in the way. In the worst case, if these hurdles cannot be overcome, our model may still provide a consistent phenomenology of the Newtonian setting. Hoping that some of its features survive the generalisation, applications of this model to specific problems of physical interest could be explored even if a fully consistent theory is still lacking. In this spirit, applications to black-holes and cosmological models of the early universe  might be tractable and should definitely be considered in the future.

\begin{acknowledgments}
We acknowledge useful discussions with Denis Bernard, Jean Bricmont, Detlef D\"urr, Sheldon Goldstein and Flavien Simon. A.T. was supported in part by Agence Nationale de la Recherche contract ANR-14-CE25-0003-01, L.D. was supported by EU COST Actions MP1006, MP1209.
\end{acknowledgments}

\appendix*

\section{Theory of monitoring and feedback}\label{monitoring_feedback}
\subsection{Continuous monitoring}
Time-continuous quantum measurement has a long history \cite{barchielli1986,caves1987,diosi1988,barchielli1991,wiseman1996,belavkin1992,diosi2006}, and has become part of standard quantum physics. It has been reviewed, e.g., in \cite{jacobs2006,barchielli2009,wiseman2009}.
One possible way to understand continuous quantum measurement is to see it as a limit of iterated unsharp measurements of a given observable $\hat A$. A well defined continuous limit is obtained when the successive measurements become infinitely unsharp but infinitely frequent. Equivalently, continuous quantum measurement can be seen as describing the interaction of a quantum system with a series of probes which are subsequently subjected to a sharp von Neumann measurement,  
the continuous limit is obtained when the interaction time goes to zero while the interaction rate goes to infinity \cite{attal2006,pellegrini2008} (see also \cite{qbm} for another derivation). 

We start with the simplest stochastic master equation (SME) \cite{diosi1988} which has become the  convenient form of most applications.
The outcome of the time-continuous measurement of the observable $\Ao$ is the following time-dependent signal:
\begin{equation}\label{signal}
A_t=\langle\Ao\rangle_t +\delta A_t. 
\end{equation}
The noise is Gaussian with zero mean and correlation
\begin{equation}\label{corr}
\Ev[\delta A_t \delta A_\tau]=\frac{1}{\gamma}\delta(t-\tau), 
\end{equation}
where $\gamma$ represents the strength of the continuous measurement.  In the formalism of continuous measurement theory, multiplicative white-noises are usually understood in the It\^o sense.
The evolution of the conditional state (i.e.: conditioned on the signal $A_t$)  
is governed by the following SME:
\begin{equation}\label{SME}
\frac{\upd\ro_t}{\upd t}=-\frac{i}{\hbar}[\Ho,\ro_t]-\frac{\gamma}{8\hbar^2}[\Ao,[\Ao,\ro_t]]+\frac{\gamma}{2\hbar^2}\Hcal[\Ao]\ro_t \delta A_t
\end{equation}
where 
\begin{equation}
\Hcal[\Ao]\ro_t=\{\Ao-\langle\Ao\rangle_t,\ro_t\}
\end{equation}
is the standard notation of \cite{wiseman2009}.
If the initial state is pure the SME will preserve its purity, hence for pure states the SME is equivalent
to the corresponding stochastic Schr\"odinger equation. On the other hand, if we average over the
detector outcomes (signal) then the average state satisfies simply the master equation (ME) without noise:
\begin{equation}\label{ME}
\frac{\upd\ro_t}{\upd t}=-\frac{i}{\hbar}[\Ho,\ro_t]-\frac{\gamma}{8\hbar^2}[\Ao,[\Ao,\ro_t]].
\end{equation}

\subsection{Feedback}
The signal can then be used to control subsequent evolution of the system via feedback.  The simplest way, called Markovian feedback, consists in applying a potential
proportional to the continuously measured value of $\Ao$, via the time-dependent additional Hamiltonian:
\begin{equation}\label{Vfb}
\Vo_t=A_t\Bo 
\end{equation}
 where $\Bo$ is another observable that can be chosen freely. This scheme amounts to a further infinitesimal unitary evolution $\exp(-i\Vo_t dt/\hbar)$ after the ``free'' evolution \eqref{SME} of the conditional state \cite{wiseman1993,diosi1994}:
 \begin{equation}\label{feedback}
\ro_t+\upd\ro_t\Rightarrow e^{-iA_t\Bo dt/\hbar}(\ro_t+\upd\ro_t)e^{iA_t\Bo dt/\hbar}
\end{equation}
where we identify  $\upd\ro_t$ by $\upd\ro_t$ in \eqref{SME}. Expanding the exponential up to second order, inserting \eqref{signal} for $A_t$ and carefully using the It\^o rule for the $\delta A_t$-dependent terms finally give the following SME for the complete evolution:
\begin{eqnarray}\label{SMEfb}
\frac{\upd\ro}{\upd t}&=&-\frac{i}{\hbar}[\Ho+\delta A\Bo,\ro]-\frac{i}{2\hbar}[\Bo,\{\Ao,\ro\}]\\
& &-\frac{\gamma}{8\hbar^2}[\Ao,[\Ao,\ro]]-\frac{1}{2\gamma\hbar^2}[\Bo,[\Bo,\ro]]
                            +\frac{\gamma}{2\hbar^2}\Hcal[\Ao]\ro \delta A\nonumber
\end{eqnarray}
Similarly to \eqref{ME}, the unconditional ME can be obtained from the above SME if we remove the noise terms containing $\delta A$. As before, the ME is linear.

\subsection{Generalization to multiple observables}
Monitoring and feedback theory can be generalized to the simultaneous monitoring of $n$ observables $A_\nu$.
In that case there are $n$ signals which satisfy the equation
\begin{equation}\label{signalmulti}
A_{\nu,t}=\langle\Ao_\nu\rangle_t +\delta A_{\nu,t}. 
\end{equation}
In the general case, the detectors associated to different $\Ao_\nu$'s can have intrinsically correlated outputs (this corresponds to initially entangled probes in the repeated interaction picture) so that the signal components can be correlated:
\begin{equation}\label{corrmulti}
\Ev[\delta A_{\nu,t} \delta A_{\mu,\tau}]=[\gamma^{-1}]_{\nu\mu}\delta(t-\tau), 
\end{equation}
where $\gamma$ is a non-negative real matrix. The SME then reads  
\begin{equation}\label{SMEmulti}
\frac{\upd\ro}{\upd t}\!=\!-\frac{i}{\hbar}[\Ho,\ro]\!-\!\frac{\gamma_{\nu\mu}}{8\hbar^2}[\Ao_\nu,[\Ao_\mu,\ro]]         \!+\!\frac{\gamma_{\nu\mu}}{2\hbar^2}\Hcal[\Ao_\nu]\ro \delta A_\mu
\end{equation}
where we use Einstein's convention of summation on repeated indices. Such an equation appeared in another similar context in \cite{qbm}.
The feedback ``potential'' \eqref{Vfb} can then be generalized as:
\begin{equation}\label{Vfbmulti}
\Vo=A_\nu\Bo_\nu 
\end{equation}
The same steps as before yield the SME for the complete evolution:
\begin{eqnarray}\label{SMEfbmulti}
\frac{\upd\ro}{\upd t}&=&-\frac{i}{\hbar}[\Ho+\delta A_\nu\Bo_\mu,\ro]--\frac{i}{2\hbar}[\Bo_\nu,\{\Ao_\nu,\ro\}]\nonumber\\
                            & & -\frac{\gamma_{\nu\mu}}{8\hbar^2}[\Ao_\nu,[\Ao_\mu,\ro]]
                            -\frac{\gamma^{-1}_{\nu\mu}}{2\hbar^2}[\Bo_\nu,[\Bo_\mu,\ro]]\nonumber\\
                            & &+\frac{\gamma_{\nu\mu}}{2\hbar^2}\Hcal[\Ao_\nu]\ro \delta A_\mu.
\end{eqnarray}
As before, the unconditional ME can be obtained if we remove the noise terms containing the $\delta A_\nu$'s. 
Finally, we note a possible rewriting when 
$\Ao_\nu\otimes\Bo_\nu=\Bo_\nu\otimes\Ao_\nu$. In that case,
the second term on the r.h.s. reduces to a Hamiltonian one: 
\begin{equation}\label{Hfb}
-\frac{i}{2\hbar}[\Bo_\nu,\{\Ao_\nu,\ro\}]=-\frac{i}{4\hbar}[\{\Ao_\nu,\Bo_\nu\},\ro].
\end{equation}

\subsection{Dictionary}
The previous results can be formally generalized to a continuous set of observables. In that case, the indices $\mu$ and $\nu$ become continuous variables $\rv$ and $\sv$ (in our case, positions in $\mathds{R}^3$) and the non-negative correlation matrix $\gamma_{\mu\nu}$ becomes a continuous kernel $\gamma_{\rv \sv}$. Finally, the dictionary to go from the notations of the appendix to the continuous setting of section \ref{monotoring_back-action} is:
\begin{eqnarray}\label{ABrhoPhi}
\nu,\mu&\Rightarrow& \rv,\sv\nonumber\\
\Ao_\nu,A_\nu,\delta A_\nu&\Rightarrow& \denso_\si(\rv),\dens(\rv),\delta\dens(\rv)\nonumber\\
\gamma_{\nu\mu}&\Rightarrow&\gamma_{\rv\sv}\nonumber\\
\Bo_\nu&\Rightarrow& \Phio_{(\si)}(\rv).
\end{eqnarray}

\bibliography{feedbackgravity}

\end{document}